\documentclass[prd,reprint,showpacs,showkeys,superscriptaddress]{revtex4}
\usepackage{amsfonts}
\usepackage{amssymb}
\usepackage{amsmath}
\usepackage{graphicx}
\usepackage[font={footnotesize,it}]{caption}
\usepackage{latexsym}
\usepackage{enumitem}
\newtheorem{theorem}{Theorem}
\begin{document}

\title{Quantum Probe of Time-like Naked Singularities for Electrically and Magnetically Charged Black Holes in a Model of Nonlinear Electrodynamics}

\author{M. Mangut}
\email{mert.mangut@emu.edu.tr}
\affiliation{Department of Physics, Faculty of Arts and Sciences, Eastern Mediterranean
University, Famagusta, North Cyprus via Mersin 10, Turkey}
\author{O. Gurtug}
\email{ozaygurtug@maltepe.edu.tr}
\affiliation{T. C. Maltepe University, Faculty of Engineering and Natural Sciences,
Istanbul -Turkey}

\begin{abstract}
The time-like naked singularities of the electrically and magnetically charged black hole solutions obtained in a model of nonlinear electrodynamics proposed by Kruglov is investigated within the framework of quantum mechanics. In view of quantum mechanics, the space-time is quantum regular provided that the time evolution of the test quantum wave packet uniquely propagates on an underlying background.  Rigorous calculations have shown that when the singularity is probed with specific quantum wave/particle modes, the quantum wave operator turns out to be essentially self-adjoint. Thus, the time evolution of the quantum wave/particle is determined uniquely. In the case of electrically charged black hole background, the unique evolution is restricted to s-wave only. For the two different magnetically charged black hole backgrounds, the time evolution is restricted to different modes for each case.
\end{abstract}

\maketitle

\section{Introduction}

In classical general relativity, spacetime singularities are known to be the
locations where timelike or null geodesics come to an abrupt halt such that
the known physical laws become invalid. This disturbing feature of
Einstein's theory of general relativity cannot be avoided and theorems of
Hawking and Penrose \cite{1} state that spacetime singularities are ubiquitous in exact
solutions of Einstein's equations. In order to save the deterministic nature
of the theory, Penrose has proposed cosmic censor hypothesis. The weak form
of the hypothesis states that singularities formed during gravitational
collapse are covered by horizon(s). However, the strong form of the hypothesis
asserts that Einstein's theory of general relativity is a deterministic
theory and therefore must be globally hyperbolic. Despite the tremendous effort,
this conjecture has not been proved yet. Some examples for the violation of the
cosmic censor hypothesis have been found. It has been demostrated in \cite{2} that
the gravitational collapse of sufficiently large collisionless gas spheroids
lead to a naked singularity; a singularity that is visible to distant
observers. Naked singularity formation is also possible in a variety of black hole solutions. Super-extremal Reissner-Nordstr\"{o}m and over-spinning Kerr black hole solutions are well-known examples.

Understanding the physics in this high curvature zone of the fabric of space-time structure requires new laws of physics. There is a strong belief among the physicists that this undesired feature of general relativity will be resolved by a consistent theory of quantum
gravity. The reason for this is the microscopic scale where these singularities are developed. At this scale, the laws of general relativity becomes invalid, and the general expectation is to implement the laws of the theory of quantum gravity. Unfortunately, such a consistent theory has not been constructed yet, and hence, alternative theories are put forward to understand the nature of singularities.

It has been known that one of the consequences of the model of nonlinear electrodynamics (NED), proposed long ago by Born and Infeld \cite{3} (BI), was to eliminate the curvature singularities that develop at the centre of the charged black holes. Since then,  there are number of NED models with interesting physical properties. The well-known regular black hole solutions within the frame of NED are given in \cite{4,5,6,7}. However, there are considerable amount of singular black hole solutions as well. Among the others, our concern in this paper is the model proposed by Kruglov \cite{8}. Our motivation to this model stems from the existence of the exact analytical solutions to Einstein-nonlinear Maxwell equations that describe both pure electrically and pure magnetically charged black holes. The pure magnetically charged solution was given by Kruglov \cite{8}, while pure electrically charged solution was obtained by Mazharimousavi and Halilsoy \cite{9}. The notable characteristic of both solutions in Kruglov's model is that unlike the BI NED model, depending on the values of parameters, the obtained solutions may admit naked singularities. Although thermodynamic properties of the obtained solutions in \cite{8,9} are studied in detail, nothing has been mentioned about the nature of the naked singularities that constitute a threat to Penrose's weak cosmic censor hypothesis. Our aim in this paper is to investigate the time-like naked singularities that appear in these solutions.

It is expected that the consistent quantum theory of gravity may heal the curvature singularities. This expectation paves the way to include the tools of quantum mechanics into the analysis of curvature singularities. One possible method to understand the singularities within the context of quantum mechanics is to probe the singularities with waves rather than point
particles. The prescription of such a method is given by the seminal work of
Wald \cite{10}, which was developed later by Horowitz and Marolf (HM) \cite{11} . In this
prescription, the problem of defining the dynamics of the scalar wave
evolution in non-globally static space-times is translated into the problem of
finding unique self-adjoint extension of the spatial part of the wave
operator. According to this sensible prescription, a space-time is quantum
mechanically regular, if the time evolution of a quantum wave packet is uniquely
determined for all times. Otherwise, a space-time is considered to be quantum mechanically
singular.

The structure of the paper is as follows. In section II, the action and the related Einstein - nonlinear Maxwell equations are given in a closed form for the considered model of NED. Then, the obtained solutions to the field equations for pure electrically and magnetically charged black hole geometries are presented together with the conditions that lead to naked singularities. In section III, the brief review of quantum singularity analysis is followed by the naked singularity analysis both in electrically and magnetically charged black hole space-times. The paper is concluded with results and discussions in section IV.

\section{Brief-Review of Electrically and Magnetically Charged Black Holes in a model of Nonlinear Electrodynamics}

In a recent paper of Kruglov \cite{8}, a model of NED has been proposed with a Lagrangian density given by

\begin{equation}
\mathcal{L}\left( \mathcal{F}\right) \mathcal{=}\frac{-\mathcal{F}}{1
-(\beta \mathcal{F})^{\gamma}}
\end{equation}

which is a generalization of his earlier model that considered $\gamma=1$ \cite{12,13,14}. Here, $\mathcal{F=}\frac{1}{4}F_{\mu \nu }F^{\mu \nu }=\frac{B^{2}-E^{2}%
}{2}$ is the Maxwell invariant in terms of electric and magnetic  fields $E$ and $B$, respectively. The parameter $\beta $ is a dimensionful positive constant, whereas the parameter $\gamma$ is a dimensionless constant which can take values $0\leqslant \gamma \leqslant 1$.  Note that linear Maxwell limit is recovered when $\beta =$ $0$. The electromagnetic field tensor $F_{\mu \nu }$ is described by   $F_{\mu \nu
}=\partial _{\mu }A_{\nu }-\partial_{\nu } A_{\mu }$, where $A=A_{\mu }dx^{\mu }$ stands for the potential one form. \\

The action in the Einstein-Nonlinear Maxwell theory is given by

\begin{equation}
I=\int d^{4}x\sqrt{-g}\left\{ \frac{R}{2\kappa ^{2}}+\mathcal{L}\left(
\mathcal{F}\right) \right\}
\end{equation}%
in which $R$ is the Ricci scalar and $\kappa ^{2}=8\pi G,$ $G$ is Newton's gravitational constant.

The line element in this model is assumed to be static spherically symmetrical and is described by,

\begin{equation}
ds^{2}=-f(r)dt^{2}+\frac{dr^{2}}{f(r)}+r^{2}d \theta ^{2}+r^{2}sin^{2}\theta d \varphi ^{2}
\end{equation}

The corresponding field equations can be derived through the variational principle. Hence, variation of the action with respect to $g^{\mu \nu }$ and vector potential $A$ gives the Einstein and Maxwell equations,

\begin{equation}
R_{\mu }^{\nu }-\frac{1}{2}R\delta _{\mu }^{\nu }=\frac{\kappa ^{2}}{4\pi }\left(
\mathcal{L}\left( \mathcal{F}\right) \delta _{\mu }^{\nu }-\mathcal{L}_{%
\mathcal{F}}\left( \mathcal{F}\right) F_{\mu \lambda }F^{\nu \lambda
}\right)
\end{equation}%
and
\begin{equation}
\partial _{\mu }\left( \mathcal{L}_{\mathcal{F}}\left( \mathcal{F}\right)
F^{\mu \nu }\right) =0
\end{equation}
respectively. Note that $\mathcal{L}_{\mathcal{F}}\left( \mathcal{F}\right) =\frac{\partial
\mathcal{L}\left( \mathcal{F}\right) }{\partial \mathcal{F}}$.

\subsection{ Electrically Charged Black Hole Solution}

Electrically charged black hole solution of the model of Kruglov corresponds to the case where $E\neq0$, but $B=0$. This problem has been solved by Mazharimousavi and Halilsoy in \cite{9}. The corresponding Einstein-nonlinear Maxwell equations are solved by assuming $\mathcal{F}=\frac{-E^{2}}{2}$ and $\gamma=1/2$. The metric function is found as

\begin{equation}
f(r)=1-\frac{2GM}{r}+\frac{2Gr^{2}}{3\alpha ^{2}}\left( 1+\frac{2q\alpha }{
r^{2}}\right) ^{3/2}-\frac{2Gq}{\alpha }\left( 1+\frac{r^{2}}{3q\alpha }
\right)
\end{equation}

in which $M$ and $q$ are mass and charge related integration constants and $\alpha^{2}=\beta/2$. The asymptotic structure of the solutions when $ r \rightarrow 0$ and $r \rightarrow \infty$ are given by
\begin{equation}
\lim_{r\rightarrow 0}f(r)=1-\frac{2Gq}{\alpha }-\frac{2G\widetilde{M}}{r}+
\frac{G}{\alpha ^{2}}\sqrt{2q\alpha }r-\frac{2G}{3\alpha ^{2}}r^{2}+\mathcal{
O}\left( r^{3}\right)
\end{equation}

\begin{equation}
\lim_{r\rightarrow \infty }f(r)=1-\frac{2GM}{r}+\frac{Gq^{2}}{r^{2}}-\frac{G\alpha q^{3}}{3r^{4}}+\mathcal{O}
\left( r^{-6}\right)
\end{equation}
in which $\widetilde{M}=M-\frac{2\sqrt{2}\left( q\alpha \right) ^{3/2}}{
3\alpha ^{2}}.$

The Ricci scalar is obtained as

\begin{equation}
R=\frac{8G}{\alpha^{2}}\left( 1+ \frac{\alpha q}{2r^{2}}-\frac{2r^{2}+3q\alpha}{2r\sqrt{r^{2}+2q\alpha}} \right)
\end{equation}

The Ricci scalar diverges at $r=0$, indicating true curvature singularity, which is covered by horizon(s). Horizon(s) of the metric  corresponds to the roots of the metric function and naked  singularity occurs when no real root exists. Our aim in this paper is to investigate the naked singularities in view of quantum mechanics. Therefore, it is important to know the values of the parameters that leads to naked singularity. Since it is hard to find exact expressions for the roots of the metric function, we analyzed them by the standard local maxima and minima method.

Extremal points of the function are determined using $\frac{df(r)}{dr}=0$. The minimum point $\left (d^{2}f/dr^{2}>0 \right ) $ is found as $r_{min}=\frac{a}{\sqrt{2}}$ under the condition $\zeta=\frac{a^{3}}{3\sqrt{2}}$ and $0<a<\sqrt{6}/3$. Note that $a=\frac{2Gq}{\alpha}$ and $\zeta=\frac{G^{3/2}M}{\alpha}$. The minimum value of the function is

\begin{equation}
f(r_{min})=1-\left(2-\sqrt{3}\right)a^{2}
\end{equation}

If $f(r_{min})>0$, no real roots exist. In this case, naked singularity arises. When this condition is fulfilled, the range for $a$ becomes $-\frac{1}{\sqrt{2-\sqrt{3}}}<a<\frac{1}{\sqrt{2-\sqrt{3}}}$. However, $r_{min}$ is obtained when $0<a<\sqrt{6}/3$. Thus, the intersection of the two inequalities imposes the condition on the parameter $a$ to be bounded to $0<a<\sqrt{6}/3$.

\subsection{ Magnetically Charged Black Hole Solution}
Pure magnetically charged black hole solution is given by Kruglov in \cite{8}. The Maxwell invariant is assumed to be $\mathcal{F=}$$\frac{B^{2}}{2}=\frac{q^{2}}{2r^{4}}$ where $q$ denotes magnetic charge. Two different solutions are obtained for the specific values of  $\gamma$.

\subsubsection{Case for $\gamma=\frac{1}{2}$}

In this case, the metric function is found to be
\begin{equation}
f(r)=1-\frac{Q}{r}\tan ^{-1}\left( q_{2}r\right)
\end{equation}%
where $Q=2Gq_{1}$, $q_{1}=\frac{q^{3/2}}{2^{3/2}\beta ^{1/4}}$ and $q_{2}=\frac{2^{1/4}}{\beta ^{1/4}
\sqrt{q}}.$ The asymptotic behaviour of the solution at $r \rightarrow 0$ and $ r \rightarrow \infty $ are given by
\begin{equation}
\lim_{r\rightarrow 0}f(r)=1-\frac{\sqrt{2}Gq}{\sqrt{\beta }}+\frac{2Gr^{2}}{
\beta }-\frac{2^{3/2}Gr^{4}}{\beta ^{3/2}q}+\mathcal{O}\left( r^{6}\right)
\end{equation}

\begin{equation}
\lim_{r\rightarrow \infty }f(r)=1-\frac{2Gm_{M}}{r}+\frac{Gq^{2}}{r^{2}}-
\frac{G\sqrt{\beta }q^{3}}{3\sqrt{2}r^{4}}-\frac{G\alpha q^{3}}{3r^{4}}+
\mathcal{O}\left( r^{-6}\right)
\end{equation}
in which $m_{M}=\frac{\pi q^{3/2}}{2^{7/4}\beta ^{1/4}}.$ The Ricci scalar for this particular case is obtained as

\begin{equation}
R=\kappa^{2} \frac{\sqrt{2\beta}q^{3}}{r^{2}\left(\sqrt{2}r^{2}+\sqrt{\beta}q\right)^{2}}
\end{equation}

The Ricci scalar diverges at $r=0$, indicating true curvature singularity. In the case of black holes, this singularity is covered by horizon(s). In order to find the condition for a naked singularity, we prefer to use graphical analysis due to the complexity in the analytical solution of the transcendental equations. In doing so,  we introduce a dimensional parameter $x=2^{1/4}r/\beta^{1/4}\sqrt{q}$ that modifies the transcendental metric function in the form

\begin{equation}
f(x)=1-\frac{a}{x}tan^{-1}(x)
\end{equation}

in which $a=\frac{\sqrt{2}Gq}{\sqrt{\beta}}$. In Figure 1, we plot (15) as a function of $x$ for different values of $a$.

\begin{figure}[htp]
\par
\label{figure}
\par
\begin{tabular}{cc}
\includegraphics[width=80mm]{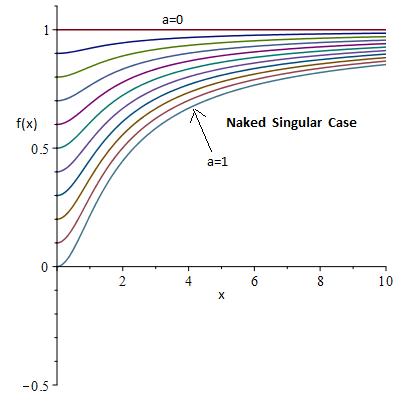} &
\end{tabular}
\caption{The transcendental metric function in terms of x with $a=0$ to $a=1$, for which the naked singularity develops.  }\centering
\end{figure}

The naked singularity zone can be found from this figure as follows: When $0<a<1$, the curves do not have an $x-$intercept, which implies the formation of the naked singularity.

\subsubsection{Case for $\gamma=\frac{3}{4}$}

The associated metric function for this particular case is found to be

\begin{equation}
f(r)=1-\frac{Gq^{3/2}}{3\sqrt{\beta}2^{3/4}r}\left[ ln\frac{l^{2}r^{2}-lr+1}{(lr+1)^{2}} + 2\sqrt{3}tan^{-1}\left( \frac{2lr-1}{\sqrt{3}} \right) \right]
\end{equation}

where $l=2^{1/4} / \beta^{1/4}\sqrt{q}$. The asymptotic structure of the solutions  near $r \rightarrow 0$ and $ r \rightarrow \infty $ are given by
\begin{equation}
\lim_{r\rightarrow 0}f(r)=1+\frac{\pi Gq^{3/2}}{3\sqrt{6}\beta^{1/4}r }-\frac{G\sqrt{q}r}{2^{1/4}\beta^{3/4}}+
\frac{2^{3/2}Gr^{4}}{5\beta ^{3/2}q}+\mathcal{O}\left( r^{6}\right)
\end{equation}

\begin{equation}
\lim_{r\rightarrow \infty }f(r)=1-\frac{2Gm_{M}}{r}+\frac{Gq^{2}}{r^{2}}-\frac{G\beta^{3/4} q^{7/2}}{2^{11/4}r^{5}}+\mathcal{O}\left( r^{-6}\right)
\end{equation}
in which $m_{M}=\frac{\pi q^{3/2}}{2^{7/2}\sqrt{3}\beta^{1/4}}$.  The Ricci scalar is calculated and given by

\begin{equation}
R=\kappa^{2} \frac{3\beta^{3/4}q^{7/2}}{4r\left(2^{3/4}r^{3}+\beta^{3/4}q^{3/2}\right)^{2}}
\end{equation}

The Ricci scalar diverges at $r=0$, and this signals the existence of the curvature singularity. When $0.21<\sqrt{\beta}/\sqrt{2}Gq$, the naked singularity develops \cite{8}.
According to the results of classical general relativity, the black hole solutions admitted by Kruglov's NED model is singular. The purpose of the present study is to investigate these classically singular space-times in view of quantum mechanics.

\subsection{ Affine Distance and Time to the Singularities }

In order to show the classical existence of singularities, one possible way is to calculate its affine paths to the singularities \cite{24} together with the required time. In doing so, we write the geodesics eqution for metric (3), which yields

\begin{equation}
\frac{\dot{r}^{2}}{f(r)}=\frac{E^{2}}{f(r)}-r^{2}\dot{\theta}^{2}+\frac{sin^{2}(\theta) l^{2}}{r^{2}}-\epsilon
\end{equation}

in which $E$ and $l$ are the conserved quantities, namely the energy and the angular momentum, respectively. The defining expressions for $E$ and $l$ are given by

\begin{equation}
\dot{t}=-\frac{E}{f(r)}, \;\;\;\; \dot{\varphi}=\frac{l}{r^{2}sin^{2}\theta}
\end{equation}

where, $\dot{}\equiv \frac{d}{d \lambda} $, $\lambda$ represents affine parameter and $\epsilon=0$ denotes null geodesics whereas $\epsilon=1$ timelike geodesics. Considering the radial null geodesics such that $l=\epsilon=0$ at the equatorial plane $( \theta=\pi/2)$, the integration of Eq.(20), gives

\begin{equation}
\lambda_{*}-\lambda_{0} \geq  \frac{1}{E}\int_{0} ^{r_{*}}dr
\end{equation}

where $\lambda_{*}$ and $r_{*}$ are finite and non-zero. Thus, the singularity for null rays is located at a finite affine distance. The time required for a photon to reach a singularity from an initial position $r_{0}$ can be calculated by using $E=-f(r)\frac{dt}{d \lambda}$, in Eq.(20) at the equatorial plane, which gives
\begin{equation}
t_{*}-t_{0} = \int_{0} ^{r_{*}} \frac{dr}{f(r)}
\end{equation}

where $t_{*}$ is the time measured by a distance observer and $t_{0}$ is the initial time.The metric function $f(r)$, for the solutions considered in this study is extremely complicated and the exact analytic solution to the above integral is not possible. The integral will be calculated for three different cases by expanding the integrand in the asymptotic form near zero, namely near the singularity, in the following subsections.

\subsubsection{Electrically Charged Solution for $\gamma=\frac{1}{2}$}

The asymptotic behavior of the integrand function near zero can be written as

\begin{equation}
 \frac{1}{f(r)}\approx-\frac{r}{a}+\frac{(d-1)r^{2}}{a^{2}}-\frac{(d-1)^{2}r^{3}}{a^{3}}+\mathcal{O}(r^{4})
\end{equation}

in which $a=2GM$, $b=\frac{2G}{3\alpha^{2}}$, $c=2q\alpha$, $d=\frac{2Gq}{\alpha}$ and $e=\frac{1}{3q\alpha}$. The calculated required time is found to be

\begin{equation}
t_{*}-t_{0} \simeq  -\frac{r_{*}^{2}}{2a}+\frac{(d-1)r_{*}^{3}}{3a^{2}}-\frac{(d-1)^{2}r_{*}^{4}}{4a^{3}}+\mathcal{O}(r_{*}^{5})
\end{equation}

\subsubsection{Magnetically Charged Solution for $ \gamma=\frac{1}{2}$ }

We can expand $1/f(r)$  near zero for calculating the integral (23). The expansion is found as

\begin{equation}
 \frac{1}{f(r)}\approx \frac{1}{1-Qq_{2}}-\frac{Qq_{2}^{3}r^{2}}{3(1-Qq_{2})^{2}}+\mathcal{O}(r^{4}).
\end{equation}

and in turn, integral (23)  is calculated and the required time to reach to the singularity is obtained as

\begin{equation}
t_{*}-t_{0} \simeq \frac{r_{*}}{1-Qq_{2}}-\frac{Qq_{2}^{3}r_{*}^{3}}{9(1-Qq_{2})^{2}}+\mathcal{O}(r_{*}^{5}).
\end{equation}

\subsubsection{Magnetically Charged Solution for $ \gamma=\frac{3}{4}$ }

The asymptotic behavior of the integrand for this particular case, near zero, can be written as

\begin{equation}
 \frac{1}{f(r)}\approx \frac{\sqrt{3}r}{\pi a}-\frac{3r^{2}}{\pi^{2}a^{2}}+\frac{3r^{3}(3\pi a^{2}l^{2}+\sqrt{3})}{\pi^{3}a^{3}}+\mathcal{O}(r^{4})
\end{equation}

in which $a=Gq^{3/2}/3\sqrt{\beta}2^{3/2}$. The time needed to reach to the singularity is found as

\begin{equation}
t_{*}-t_{0} \simeq \frac{\sqrt{3}r_{*}^{2}}{2\pi a}-\frac{r_{*}^{3}}{\pi^{2}a^{2}}+\frac{3r_{*}^{4}(3\pi a^{2}l^{2}+\sqrt{3})}{4\pi^{3}a^{3}}+\mathcal{O}(r_{*}^{5})
\end{equation}

In summary, our analysis has shown that the naked singularities in the considered spacetimes are true curvature singularities.

\section{Quantum Regularity Analysis of Time-like Naked Singularity}

To probe the singularity with quantum wave packets/particles, we use the HM prescription which incorporates with the propagation of a scalar field in the background of a classically singular space-time.  The brief review of this prescription is as follows (See references \cite{15,16,17,18,19} for a detailed mathematical background ). The massive Klein - Gordon
equation is the governing equation for the scalar field and can be written
by splitting temporal and spatial parts as
\begin{equation}
\frac{\partial ^{2}\Psi }{\partial t^{2}}=-A\Psi
\end{equation}%
in which $A=-VD^{i}\left( VD_{i}\right)+V^{2}m^{2} $ is the spatial operator with $
V^{2}=-\xi _{\mu }\xi ^{\mu }$ and $m$ denotes mass. In this setting, $\xi ^{\mu }$ denotes
timelike Killing field and $D_{i}$ is the spatial covariant derivative on a
static slice $\Sigma .$ The appropriate function space is the usual Hilbert
space $\mathcal{H}$ $=L^{2}\left( \Sigma ,V^{-1}d\sigma \right) ,$ where $%
d\sigma $ stands for the induced volume form of $\Sigma .$ If we define the
initial domain of $A$ to be $C_{0}^{\infty }\left( \Sigma \right) $ (smooth
functions of compact support on $\Sigma $), then $A$ is a real, positive,
symmetric operator and self-adjoint extensions $A_{E}$ always exist. The key
point in this prescription is to show that the self-adjoint extension $A_{E}$
is unique, which is said to be essentially self-adjoint. If this is the case, then
the Klein-Gordon equation for a free relativistic scalar particle satisfies%
\begin{equation}
i\frac{d\Psi }{dt}=\sqrt{A_{E}}\Psi
\end{equation}%
whose solution is
\begin{equation}
\Psi \left( t\right) =\exp (-it\sqrt{A_{E}})\Psi \left( 0\right)
\end{equation}%
If the extension is not unique, then the future time evolution of the wave
function is ambiguous and the space-time is said to be quantum singular.

In order to determine the number of self-adjoint extensions, the concept of deficiency indices
discovered by Weyl \cite{16} and generalized by von Neumann is used \cite{17}. The deficiency subspaces $N_{\pm}$ are defined by

\begin{equation}
\begin{aligned}
N_{+}=\{ \psi \in D(A^{*}) | \; A^{*}\psi=\lambda_{+} \psi \} \\
N_{-}=\{ \psi \in D(A^{*}) | \; A^{*}\psi=\lambda_{-} \psi \}
\end{aligned}
\end{equation}

where $\lambda_{+}=i\lambda$ and  $\lambda_{-}=-i\lambda$, with $\lambda$ representing an arbitrary positive constant. The dimensions, $n_{+}=dim[ N_{+ }]$ and $n_{-}=dim[ N_{-} ]$ represent the deficiency indices of the operator $A$. The determination of deficiency indices is then reduced to counting the number of solutions of $A^{*}\psi= \mp i \lambda \psi$, for $\lambda=1$,

\begin{equation}
( A^{*} \mp i ) \psi =0
\end{equation}

that belong to the Hilbert space $\mathcal{H}$. Those solutions which are not square integrable do not belong to the Hilbert space, and hence, the deficiency indices $n_{+}=n_{-}=0$. The related theorem \cite{16,17}, which will be used to define whether the self-adjoint extension of the operator $A$ is unique or not, is as follows.

\begin{theorem}
For an operator $A$ with deficiency indices $(n_{+},n_{-})$, there are three possibilities

(i) If $n_{+}=n_{-}=0$,  then $A$ is (essentially) self-adjoint (in fact, this is a necessary and sufficient condition).

(ii) If $n_{+}=n_{-}=n \geq 1$ , then $A$ has infinitely many self-adjoint extensions, parametrized by a unitary $n \times n$ matrix.

(iii) If $n_{+} \neq n_{-}$, then $A$ has no self adjoint extension.

\end{theorem}

Based on this theorem, if there are no square integrable solutions for each sign of Eq.(34) for all space $(0,\infty)$, operator $A$ possesses a unique self-adjoint extension and thus, it is essentially self-adjoint.

\subsection{Massive Scalar Fields}

The massive Klein-Gordon equation in general is given by

\begin{equation}
\left [ \frac{1}{\sqrt{-g}} \partial_{\mu}[\sqrt{-g}g^{\mu \nu}\partial_{\nu}]-m^{2}\right]\psi=0
\end{equation}

in which $m$ is the mass of the scalar particle. The Klein-Gordon equation is written for metric (3) and after
separating time and spatial parts, we find

\begin{equation}
\frac{\partial ^{2} \psi}{\partial t^{2}}=f^{2}(r)\frac{\partial ^{2} \psi}{\partial r^{2}}+\frac{f(r)}{r^{2}}\frac{\partial ^{2} \psi}{\partial \theta^{2}}+\frac{f(r)}{r^{2}sin^{2}\theta}\frac{\partial ^{2} \psi}{\partial \varphi^{2}}+\frac{f(r)cot \theta}{r^{2}}\frac{\partial  \psi}{\partial \theta}+f(r)\left ( \frac{2f(r)}{r}+f^{'}(r)\right)\frac{\partial  \psi}{\partial r}-f(r)m^{2}\psi
\end{equation}

When we compare equations Eq.(36) and Eq.(30), the spatial part of the wave operator can be written as

\begin{equation}
A=-f^{2}(r)\frac{\partial ^{2} }{\partial r^{2}}-\frac{f(r)}{r^{2}}\frac{\partial ^{2} }{\partial \theta^{2}}-\frac{f(r)}{r^{2}sin^{2}\theta}\frac{\partial ^{2} }{\partial \varphi^{2}}-\frac{f(r)cot \theta}{r^{2}}\frac{\partial }{\partial \theta}-f(r)\left ( \frac{2f(r)}{r}-f^{'}(r)\right)\frac{\partial}{\partial r}+f(r)m^{2}
\end{equation}

Using separation of variables for Eq.(34), in the form of  $ \psi = R(r)Y_{l}^{m}( \theta , \varphi )$,  yields the following radial  part  for $R(r)$

\begin{equation}
R^{''}+\frac{[r^{2}f(r)]^{'}}{r^{2}f(r)}  R^{'}+\left [ \frac{-l(l+1)}{r^{2}f(r)}-\frac{  m^{2}}{f(r)} \mp \frac{i}{f^{2}(r)}\right]R=0
\end{equation}

where prime denotes the derivative with respect to $r$ and $R = R(r)$.

The square integrability condition of the solutions (38) for each sign $ \pm$ is verified by calculating the squared norm, in which the function space on each $t=constant$  hypersurface $\Sigma _{t}$ is defined as $\mathcal{H}=\{ R : ||R||<\infty \}$. The squared norm for $(3+1)-$ dimensional space is written by \cite{11}

\begin{equation}
||R||^{2}=\int_{\Sigma _{t}}\sqrt{-g}g^{tt}RR^{*}d^{3}\Sigma _{t}.
\end{equation}

The spatial wave operator $A$ is essentially self-adjoint  if  the solutions of  Eq.(38) are not square integrable over all Hilbert space  $\mathcal{H}$.

\subsection {Quantum Probe of Electrically Charged Black Hole Solution }
The asymptotic behavior of Eq.(38), near $r \rightarrow 0$ and $r \rightarrow  \infty$, will be studied  separately in the following subsections.

\subsubsection{The  case of $r \rightarrow 0$}

The approximate metric function when $ r \rightarrow \ 0 $  is written as

\begin{equation}
f(r) \approx \sigma-\frac{\beta}{r}+\mathcal{O}\left( r\right) \label{ilk}
\end{equation}

in which $\sigma=1-\frac{2Gq}{\alpha}$ and $\beta=2GM-\frac{4G\sqrt{2}(q\alpha)^{3/2}}{3 \alpha ^{2}}$. Thus, Eq.(38) reduces to

\begin{equation}
R^{''}+ \frac{1}{r}R^{'}+\rho R =0
\end{equation}

where $\rho=\frac{l(l+1)}{\beta}$. The solution is found in the form

\begin{equation}
R(r)=C_{1}J_{0}(2\sqrt{\rho r})+C_{2}Y_{0}(2\sqrt{\rho r})
\end{equation}

in which $C_{1},C_{2}$ are arbitrary constants and $J_{0}$ and $Y_{0}$ are the Bessel function of the first kind of order zero and the Bessel function of the second kind of order zero. The behaviour of the Bessel functions for real $\nu \geq 0$ as $ r \rightarrow 0 $ are defined by \cite{22}

\begin{equation}
\begin{aligned}
J_{\nu}(r) &\sim \frac{1}{\Gamma(\nu +1)}\left ( \frac{r}{2}\right )^{\nu}\\
Y_{\nu}(r) &\sim \begin{cases}
\frac{2}{\pi}\left[ ln \left ( \frac{r}{2} \right ) +\gamma \right ]&,\nu =0 \;and \; \gamma \cong 0.5772\\
-\frac{\Gamma(\nu)}{\pi}\left(\frac{2}{r}\right)^{\nu}&,\nu\neq 0 \: \: \: \: \: \: \: \: \:\: \: \: \: \: \: \: \: \: \: \: \: \: \: \: \: \: \: \: \: \: \: \:
\end{cases}
\end{aligned}
\end{equation}

Thus, the solution of Eq.(42) can be written as

\begin{equation}
R(r) \cong k_{1}+ k_{2} ln(\rho r)
\end{equation}

with $k_{1}= C_{1}+\frac{2C_{2}\gamma}{\pi}$ and $k_{2}=\frac{C_{2}}{\pi}$. Once Eq.(44) is substituted into the squared norm (39),  it becomes

\begin{equation}
\| R \|^{2} \sim  \int_{const.}^{0}\frac{r^{3}|R|^{2}}{r\sigma-\beta}dr=k_{1}^{2}\int_{const.}^{0}\frac{r^{3}}{r\sigma-\beta}dr+2k_{1}k_{2}\int_{const.}^{0}\frac{r^{3}ln(\rho r)}{r\sigma-\beta}dr+k_{2}^{2}\int_{const.}^{0}\frac{r^{3}ln^{2}(\rho r)}{r\sigma-\beta}dr.
\end{equation}

The first integral gives

\begin{equation}
I_{1}=k_{1}^{2}\int_{const.}^{0}\frac{r^{3}}{r\sigma-\beta}dr=k_{1}^{2} \left \{\frac{\beta^{3}ln(\beta-\sigma r)}{\sigma^{4}}+\frac{\beta^{2}r}{\sigma^{3}}+\frac{\beta r^{2}}{2\sigma^{2}}+\frac{r^{3}}{3\sigma} \right \} < \infty
\end{equation}

which belongs to Hilbert space. The convergence character of second integral is analyzed by using the comparison test. The following inequality can be defined as a requirement of the comparison test,

\begin{equation}
0 \leq \frac{r^{3}ln(\rho r)}{r \sigma - \beta} \leq \frac{ln(\rho r)}{r \sigma - \beta}
\end{equation}

where $r>0$ and $r<<1$. The integral of $ln(\rho r)/(r \sigma - \beta)$ can be calculated by

\begin{equation}
\int_{const.}^{0} \frac{ln(\rho r)}{r \sigma - \beta}dr=\frac{ln(\rho r)ln(1-\frac{\sigma r}{\beta})+ Li_{2} (\frac{\sigma r}{\beta})}{\sigma} |_{const.}^{0}
\end{equation}

in which $Li_{2}(r)$ is the polylogarithm function. Since $\lim_{r\to 0}=ln(\rho r)ln(1-\frac{\sigma r}{\beta}) =0$  and $\lim_{r\to 0}Li_{2}(r)=0$ integral (48) converges. According to the comparison test, the convergence of integral $\int_{const.}^{0}\left( \frac{ln(\rho r)}{r \sigma - \beta}\right)dr$,  implies the convergence of $\int_{const.}^{0} \frac{r^{3}ln(\rho r)}{r \sigma - \beta}dr$. If we apply similar analysis to the third integral, the inequality can be written as

\begin{equation}
0 \leq \frac{r^{3}ln^{2}(\rho r)}{r \sigma - \beta} \leq \frac{ln^{2}(\rho r)}{r \sigma - \beta}
\end{equation}

The integral of $\frac{ln^{2}(\rho r)}{r \sigma - \beta}$ can be found as

\begin{equation}
\int_{const.}^{0} \frac{ln^{2}(\rho r)}{r \sigma - \beta}dr=\frac{ln^{2}(\rho r)ln(1-\frac{\sigma r}{\beta})+2ln(\rho r)Li_{2}(\frac{\sigma r}{\beta})- Li_{3} (\frac{\sigma r}{\beta})}{\sigma} |_{const.}^{0}
\end{equation}

where $Li_{2}(r)$ and $Li_{3}(r)$  are the polylogarithm function. As $\lim_{r\to 0}=ln(\rho r)^{2}ln(1-\frac{\sigma r}{\beta}) =0$,     $\lim_{r\to 0}ln(\rho r)Li_{2}(\frac{\sigma r}{\beta})$ and $\lim_{r\to 0}Li_{3}(\frac{\sigma r}{\beta})=0$ the integral (50) is convergent. According to the comparison test, convergence of integral $\int_{const.}^{0}\left( \frac{ln^{2}(\rho r)}{r \sigma - \beta}\right)dr$,  implies the convergence of $\int_{const.}^{0} \frac{r^{3}ln^{2}(\rho r)}{r \sigma - \beta}dr$. Thus, whenever the constant parameter satisfies $\rho\neq0$, all the solutions belong to Hilbert space. However, for the $s-$wave mode only, which is the case when $\rho=0$, the square integrability condition states that $||R||^{2} \rightarrow \infty$ . In this case, the solution no longer belongs to Hilbert space.\\

At this stage, the physics of this behavior deserves more critical explanations. In fact, the essential self - adjointness of the Hamiltonian wave operator can be understood with two different methods. One of the method is the von Neumann deficiency indices, which is used in this study. The other method is the Weyl's \textit{limit circle - limit point} criterion that relates the essential self - adjointness of the Hamiltonian operator to the behavior of the effective potential of the one - dimensional Schr\"{o}dinger - like wave equation that determines the behavior of the wave packet. This behavior is explained very well in \cite{23}. What is exactly stated is that the effective potential is sufficiently repulsive at the origin if it is in limit-point case. The condition for this particular case is verified (via Theorem X.10 of Reference \cite{19}) in \cite{23} as follows: If the effective potential $ V(x)\rightarrow C_{0}x^{-n}$ as $ x\rightarrow 0 $, here $C_{0}$ is a constant parameter. For an effective potential to be in the limit-point case, which in turn determines the repulsive nature, either $n>2$ or $n=2$  (together with $C_{0}\geq\frac{3}{4}$) should be satisfied. Note that for both conditions, it is required to have $C_{0}\geq0$.\\

In view of this definition, for the case when $l=0$, the effective potential is sufficiently repulsive to enable the healing of quantum singularity, whereas for higher wave modes with $l \geq 1 $, the centrifugal term also comes into play and the effective potential no longer seems to be able to possess the sufficient repulsive effect, and hence, remains quantum singular. This effect can be recognised via the following analysis.\\

From Eq. (\ref{ilk}), the Schr\"{o}dinger-like form of the Klein-Gordon equation (35) for $r \rightarrow 0$  can be expressed as

\begin{equation}
\frac{d^2 R}{d\tilde{r_*}^2}+\left\{\tilde{\omega}^{2}-\left[\frac{\beta^2}{\left(\tilde{r_*} \bar{\beta}\right)^{2}}+\frac{l(l+1)-\sigma\beta}{\left(\tilde{r_*}\bar{\beta}\right)^{3/2}}+\sigma m^{2} - \frac{\beta}{\sqrt{\tilde{r_*}\bar{\beta}}}m^{2}\right]\right\}R=0
\end{equation}

where $\tilde{\omega}$ represents the wave frequency, $\tilde{r_*}=r_{*}-\frac{\beta (\ln{\beta}-1)}{\sigma^2}$ and $\bar{\beta}=\frac{4G\sqrt{2}(q\alpha)^{3/2}}{3 \alpha ^{2}}-2GM$, i.e. $\bar{\beta}=-\beta$. Note that $r_{*}$ stands for the tortoise coordinate which has been evaluated via the definition $r_{*}=\int \frac{dr}{f(r)}$, for $r \rightarrow 0$. Thus, one can also state $\tilde{r_*}\cong r^{2}/\bar{\beta}$. In brief, the effective potential for the s-wave mode takes the form $V_{eff}=1/\tilde{r_*}^2$ which implies that its coefficient $1 > \frac{3}{4}$ and thus satisfies the limit - point condition. This result displays the existence of the repulsive potential that shields the singularity for the s-wave probe.

\subsubsection { The  case of $ r \rightarrow \infty $}

 The approximate metric function when $ r \rightarrow \infty $  is found by

\begin{equation}
f(r) \approx 1-\frac{\delta}{r}+\mathcal{O}\left( r^{-2}\right)
\end{equation}

where $\delta=2GM$. Eq.(38) becomes

\begin{equation}
R^{''}+ \frac{2}{r}R^{'}+[-m^{2} \mp i ]R =0
\end{equation}

whose solution is

\begin{equation}
R(r)=\frac{C_{3}}{r}sin \omega r+\frac{C_{4}}{r}cos \omega r
\end{equation}

in which  $ \omega = \sqrt{- m^{2} \mp i}$  and $C_{3},C_{4}$  are  the  integration  constants. Substituting Eq.(54)  into Eq. (39) leads to

\begin{equation}
\| R \|^{2} \sim  \int_{const.}^{\infty}\frac{r^{3}|R|^{2}}{r-\delta}dr
\end{equation}

If $C_{3}=C_{4}=1$ , Eq.(55) can be written as

\begin{equation}
\| R \|^{2} \sim  \int_{const.}^{\infty}\frac{r}{r-\delta}(1+2sin(\omega r) cos(\omega r))dr= \int_{const.}^{\infty}\frac{r}{r-\delta}dr+2 \int_{const.}^{\infty}\frac{rsin(\omega r) cos(\omega r)}{r-\delta}dr
\end{equation}

The first integral is easy to integrate and gives

\begin{equation}
\int_{const.}^{\infty}\frac{r}{r-\delta}dr=(r-\delta+\delta ln|r-\delta|) |_{const.}^{\infty}\rightarrow\infty
\end{equation}

The second integral is analyzed by using the comparison test as in the previous case.  The second integral can be written as

\begin{equation}
I=\int_{const.}^{\infty}\frac{rsin(2\omega r)}{r-\delta}dr
\end{equation}

Substituting

\begin{equation}
sin(2\omega r)=\sum_{n=0}^{\infty} (-1)^{n} \frac{(2\omega r)^{2n+1}}{(2n+1)!}
\end{equation}

into Eq.(58) gives

\begin{equation}
I=\int_{const.}^{\infty}\left(\frac{r}{r-\delta}\right) \left\{\sum_{n=0}^{\infty} (-1)^{n} \frac{(2\omega r)^{2n+1}}{(2n+1)!}\right\} dr=\left\{\sum_{n=0}^{\infty} (-1)^{n} \frac{(2\omega )^{2n+1}}{(2n+1)!}\right\}\int_{const.}^{\infty}\left(\frac{r^{\zeta}}{r-\delta}\right)dr
\end{equation}

in which  $\zeta = 2n+2$. It should be noted that the series in front of the second integral is analysed with D'Alambert ratio test for  convergency. It is found that the series is absolute convergent.  Letting $z=r-\delta$ and noting $r>>1$ implies $z>>1$. Then, the second integral becomes proportional to

\begin{equation}
\sim  \int_{const.}^{\infty}\frac{(z+\delta)^{\zeta}}{z}dz.
\end{equation}

By using the comparison test, we can define the following inequality

\begin{equation}
0 \leq \frac{z+\delta}{z} \leq \frac{(z+\delta)^{\zeta}}{z}.
\end{equation}

The  integral of $\frac{z+\delta}{z}$ can be calculated easily and it is found that it diverges

\begin{equation}
\int_{const.}^{\infty}\left( \frac{z+\delta}{z}\right)dz=(z+\delta ln|z|) |_{const.}^{\infty}\rightarrow\infty
\end{equation}

According to the comparison test, divergence of integral $\int_{const.}^{\infty}\left( \frac{z+\delta}{z}\right)dz$  implies the divergence of $\int_{const.}^{\infty} \frac{(z+\delta)^{\zeta}}{z}dz$. As a result of this analysis, the solution  (54) fails to satisfy the square integrability condition. As a consequence, it does not belong to Hilbert space.

All these calculations have shown that when the singularity is probed with quantum $s$-wave mode, the spatial wave operator is essentially self-adjoint and hence, the time evolution of the quantum wave packet is uniquely determined. This indicates that the classical naked singularity in the electrically charged black hole geometry is quantum mechanically regular for this specific choice of mode. Otherwise, it is quantum singular.

\subsection {Quantum Probe of  Magnetically Charged Black Hole Solution  }

Kruglov \cite{8} has obtained two different magnetically charged black hole solutions by taking two different values of $\gamma$, namely $\gamma=1/2$ and $\gamma=3/4$. Formation of naked singularities in each case will be probed with scalar waves obeying the Klein-Gordon equation.

\subsubsection{The Case of $\gamma=1/2$}

The approximate metric function when $ r \rightarrow \ 0 $  is written as

\begin{equation}
f(r) \approx \eta+\mathcal{O}\left( r^{2}\right) \label{E2}
\end{equation}

in which $\eta=1-\frac{\sqrt{2}Gq}{\sqrt{\beta}}$. Thus, Eq.(38) become

\begin{equation}
R^{''}+ \frac{2}{r}R^{'}-\frac{l(l+1)}{\eta r^{2}} R =0.
\end{equation}

The solution of this equation is given by

\begin{equation}
R(r)=a_{1}r^{\frac{(-1+\sqrt{1+4c})}{2}}+a_{2}r^{\frac{(-1-\sqrt{1+4c})}{2}}
\end{equation}

where $c=\frac{l(l+1)}{\eta}$ and $a_{1},a_{2}$  are  the  integration  constants. Substituting Eq.(66)  into Eq. (39) yields

\begin{equation}
\| R \|^{2} \sim  \int_{const.}^{0}\frac{r^{2}|R|^{2}}{\eta}dr=a_{1}^{2}\int_{const.}^{0}\frac{r^{1+t}}{\eta}dr+2a_{1}a_{2}\int_{const.}^{0}\frac{1}{\eta}dr+a_{2}^{2}\int_{const.}^{0}\frac{r^{1-t}}{\eta}dr
\end{equation}

in which $t=\sqrt{1+4c}$. The square integrability analysis has revealed that whenever $3/4<l(l+1)/\eta$, the squared norm diverges. This implies that the solution for this particular choice fails to be square integrable. However, if $3/4>l(l+1)/\eta$ then the solution is square integrable. The square-integrability can also be noticed via analysing the behaviour of the effective potential when $r \rightarrow 0$. With this purpose, Eq.(35) together with the Eq.(64) can be written in the form of one - dimensional Schr\"odinger - like wave equation as
\begin{equation}
\frac{d^2 R}{d{r_*}^2}+\left\{\tilde{\omega}^{2}-\left[\frac{l(l+1)}{\eta r_{*}^{2}}+\eta m^{2}\right]\right\}R=0.
\end{equation}
In this case, the tortoise coordinate reads  $r_{*}=\eta^{-1} r$. Ultimately, the s-wave mode experiences an attractive effective potential which goes as $V_{eff}=\eta m^2$; and hence one cannot heal the quantum singularity. From the associated effective potential, it can be seen that for $l\neq0$, the solution will be quantum regular if and only if  $3/4<l(l+1)/\eta$, which also agrees with our calculations obtained from the von Neumann deficiency indices.\\

The approximate metric function when $ r \rightarrow \infty $  is found by

\begin{equation}
f(r) \approx 1-\frac{\epsilon}{r}+\mathcal{O}\left( r^{-2}\right)
\end{equation}

where $\epsilon=2Gm_{M}$. Eq.(38) takes the form

\begin{equation}
R^{''}+ \frac{2}{r}R^{'}+[-m^{2} \mp i ]R =0.
\end{equation}

The solution of this equation is written as

\begin{equation}
R(r)=\frac{a_{3}}{r}sin \omega r+\frac{a_{4}}{r}cos \omega r
\end{equation}

in which  $ \omega = \sqrt{- m^{2} \mp i}$  and $a_{3},a_{4}$  are  the  integration  constants. The square integrability analysis for this case is very similar to the one performed for electrically charged black hole. Applying the same steps for integration indicates that $||R||^{2} \rightarrow \infty$ , hence the solution does not belong to Hilbert space.

As a result, quantum probe of the naked singularity in magnetically charged black hole for $\gamma=1/2$, has shown that as long as $3/4<l(l+1)/\eta$, the spatial part of the wave operator is essentially self-adjoint and the time evolution is uniquely determined for all times. This shows that the classically singular space-time becomes quantum mechanically regular for the specific mode of scalar waves obeying the Klein-Gordon equation.

\subsubsection{The Case of $\gamma=3/4$}

The approximate metric function when $ r \rightarrow \ 0 $  is given by

\begin{equation}
f(r) \approx 1+\frac{\lambda}{r}+\mathcal{O}\left( r\right) \label{E3}
\end{equation}

in which $\lambda=\frac{\pi G q^{3/4}}{3\sqrt{6}\beta^{1/4}}$. Thus, Eq.(38) reduces to

\begin{equation}
R^{''}+ \frac{1}{r}R^{'}-\tilde{\lambda} R =0
\end{equation}

where $\tilde{\lambda}=\frac{l(l+1)}{\lambda}$. The solution is found by

\begin{equation}
R(r)=b_{1}I_{0} \left (2\sqrt{\tilde{\lambda} r}\right)+b_{2}K_{0} \left(2\sqrt{\tilde{\lambda} r}\right)
\end{equation}

where $b_{1},b_{2}$ are arbitrary constants and $I_{0}$ and $K_{0}$ are the modified Bessel functions of the first kind of order zero and the modified Bessel functions of the second kind of order zero. The behaviour of the modified Bessel functions for real $\nu \geq 0$ as $ r \rightarrow 0 $ are given by \cite{22}

\begin{equation}
\begin{aligned}
I_{\nu}(r) &\sim \frac{1}{\Gamma(\nu +1)}\left ( \frac{r}{2}\right )^{\nu}\\
K_{\nu}(r) &\sim \begin{cases}
-lnz &,\nu =0 \\
\frac{\Gamma(\nu)}{2}\left(\frac{2}{r}\right)^{-\nu}&,\mathfrak{R}(\nu)>0 \: \: \: \: \: \: \: \: \:\: \: \: \: \: \: \: \: \: \: \: \: \: \: \: \: \: \: \: \: \: \: \:
\end{cases}
\end{aligned}
\end{equation}

Thus, the solution of Eq.(74) can be written as

\begin{equation}
R(r) \cong b_{1}-\frac{b_{2}}{2} ln(4\tilde{\lambda} r)
\end{equation}

All probes, except $s-$waves, satisfy the square integrability condition. As already stated, one can double-check whether our results make sense intuitively by observing the behavior of the effective potential at the origin. Klein-Gordon Equation (35) can be presented in Schr\"odinger-like form by using Eq. (\ref{E3}) as

\begin{equation}
\frac{d^2 R}{d\tilde{r_*}^2}+\left\{\tilde{\omega}^{2}-\left[\frac{\lambda^2}{\left(\tilde{r_*} \lambda\right)^{2}}+\frac{l(l+1)+\lambda}{\left(\tilde{r_*}\lambda\right)^{3/2}}+ m^{2} + \frac{\lambda}{\sqrt{\tilde{r_*}\lambda}}m^{2}\right]\right\}R=0
\end{equation}

where $\tilde{r_*}=r_{*}-\lambda (\ln{\lambda}-1)$ and the tortoise coordinate is $\tilde{r_*}\cong r^{2}/\lambda$. Then, just like the electrical case, we will be having $V_{eff}=1/\tilde{r_*}^2$ which again will imply quantum non-singularity for the s-wave mode.\\

The approximate metric function when $ r \rightarrow \infty $  is given by

\begin{equation}
f(r) \approx 1-\frac{\tilde{\delta}}{r}+\mathcal{O}\left( r^{-2}\right)
\end{equation}

where $\tilde{\delta}=2Gm_{M}$. Eq.(38) reduces

\begin{equation}
R^{''}+ \frac{2}{r}R^{'}+[-m^{2} \mp i ]R =0
\end{equation}

whose solution is found to be

\begin{equation}
R(r)=\frac{b_{3}}{r}sin \omega r+\frac{b_{4}}{r}cos \omega r
\end{equation}

in which  $ \omega = \sqrt{- m^{2} \mp i}$  and $b_{3},b_{4}$  are  the  integration  constants.  The analysis of square integrability is similar to the former analysis, because of the similarity of the solution (80). The squared norm of this solution diverges, thus the solution does not belong to Hilbert space.

Quantum singularity analysis of the naked singularities that emerges in the magnetically charged black hole solutions of Kruglov's NED model is shown to be quantum regular with respect to the restricted mode of wave probe. Quantum regularity is possible if $s$-mode waves are used, otherwise the space-time remains quantum singular.

\section{Results and Discussions}

Understanding and resolving the space-time singularities in classical general relativity may constitute one of the most challenging subjects. Despite the fact that the space-time singularities are predictions of Einstein's theory of general relativity, the theory itself  cannot explain the singularities and the interesting thing it becomes invalid at the scales where the singularities are develop. As a result of this fact, a new physics with  new tools are required to avoid these singularities in the fabric of space-time structure.

In this paper, we have investigated the formation of time-like naked singularities that appears in a model of NED, proposed by Kruglov. The main motivation behind the choice of this model is the existence of exact electrically and magnetically charged black hole  solutions to Einstein - nonlinear Maxwell equations. In contrast to the formerly obtained solutions in Einstein coupled to NED, Kruglov's model may admit singularities, which are not covered by horizon(s), and thus, forms a threat to cosmic censor hypothesis.  Our analysis of naked singularities incorporates with quantum mechanics in such a way that the singularities are probed with quantum wave/particle obeying the Klein-Gordon equation. The key idea in the analysis is to check whether the quantum operator is essentially self-adjoint or not.

It has been shown by rigorous calculations that the electrically charged black hole solution possessing $\gamma=1/2$ becomes quantum mechanically regular for the case when the singularity is probed with s-waves. Otherwise, the space-time remains quantum singular. In addition to the electrical case, two different magnetically charged solutions corresponding to $\gamma=1/2$ and $\gamma=3/4$ are also investigated separately. The analysis has shown that when $\gamma=1/2$, the space-time remains quantum regular provided that the mode of the wave is $3/4<l(l+1)/\eta$. Otherwise, the space-time becomes quantum singular.  Our results have shown that the case for $\gamma=3/4$ is very similar to the electrically charged case. The time evolution of the quantum wave is uniquely determined only if s-wave mode is used.

The behavior of the naked singularity against a quantum particle/wave probe both in electrically and magnetically charged cases are analysed in detail to figure out the underlying physics. It has been known that the time-like naked singularity of the over - extreme Reissner - Nordstr\"{o}m space-time is quantum mechanically singular even if it is probed with s-waves \cite{15}. The main reason of this is hidden in the effective potential. Intuitively speaking, the effective potential is not sufficiently repulsive to form a barrier. This reality becomes apparent, if the Hamiltonian wave operator for the over - extreme Reissner - Nordstr\"{o}m  space-time is written near the singularity, namely as $r\rightarrow0$. In this limit, the corresponding metric function becomes $f(r)\cong\frac{Q}{r^2}$ and the tortoise coordinates reads $r_*\cong(r^{3}/3Q)$ for $r\rightarrow0$. Hence, the one - dimensional Schr\"{o}dinger - like wave equation becomes

\begin{equation}
\frac{d^2 R}{dr_{*}^2}+\left\{\tilde{\omega}^{2}-\left[\frac{2}{9r_{*}^{2}}+\frac{l(l+1)Q}{(3r_* Q)^{4/3}}+\frac{Qm^{2}}{(3Qr_*)^{2/3}}\right]\right\}R=0.
\end{equation}

What we observe from this equation is that irrespective of the modes of waves, the potential $ \frac{2}{9r_{*}^{2}} $ is not sufficiently repulsive (or not in the limit-point case), as the coefficient  $C_{0}=\frac{2}{9}<\frac{3}{4}$. However, in the considered NED model, the calculated effective potentials indicate that the electrically ($\gamma=1/2$) and magnetically ($\gamma=3/4$) charged cases obey potentials of the same form. Therefore, it is very natural and expected to observe similar characteristics in the quantum singularity structure. Consequently, the time-like naked singularities in both cases become quantum regular if and only if s-wave mode is considered.

In summary, the quantum singularity analysis of the considered NED model has shown that for \textit{specific}  modes of wave probe, the appeared naked singularities are removed for both the electrically and magnetically charged black hole geometries. However, in the generic case, the space-time remains quantum singular. This happens if bosonic (spin-0) waves/particles are used.

As a future research direction of this analysis, wave probes can be extended to spinorial fields, namely, electromagnetic (spin-1) and Dirac (spin-1/2) fields to see whether the spin of the wave is effective in healing the singularities or not.

\end{document}